\documentclass[sigconf,nonacm]{acmart} % anonymous

\AtBeginDocument{%
  }

\setcopyright{none}
\acmConference{}{}{}
\acmBooktitle{Gen-IR@SIGIR24 -- Second Workshop on Generative Information Retrieval -- SIGIR'24,
 July 17, 2024, Washington, DC}
\usepackage[T1]{fontenc}
\usepackage{graphicx}
\usepackage{dsfont}
\usepackage{color}
\usepackage{makecell}
\usepackage{listings} % code blocks
\usepackage{caption}
\DeclareCaptionType{code}[Code Listing][List of Codes]
\newenvironment{codeblock}{\captionsetup{type=code}}{}
\DeclareCaptionLabelFormat{andtable}{#1~#2  \&  \tablename~\thetable}
\usepackage{pifont}% http://ctan.org/pkg/pifont
\newcommand{\xmark}{\ding{55}}%
\newcommand{\cmark}{\ding{51}}%

\newcommand{\popa}{\phi(a)}
\newcommand{\popai}{\phi(a_i)}
\newcommand{\gapr}{AP(r)}
\newcommand{\gapu}{AP(u)}
\newcommand{\gagpr}{AggP(r)}
\newcommand{\gagpu}{AggP(u)}

\newcommand{\ita}{a}
\newcommand{\profu}{p_u}
\newcommand{\profr}{p_r}
\newcommand{\deltgaru}{\Delta AP(r,u)}
\newcommand{\deltgaur}{\Delta AP(u,r)}

\newcommand{\deltaru}{M(r,u)}
\newcommand{\deltaur}{M(u,r)}

\begin{document}

\title{Large Language Models as Recommender Systems: A Study of Popularity Bias}
%
% \titlerunning{LLMs and Recommender Systems: A Study of Popularity Bias}

\author{Jan Malte Lichtenberg}
\authornote{The authors contributed equally to this research.}
\email{jlichten@amazon.de}
\orcid{0000-0002-7140-2520}
\affiliation{%
  \institution{Amazon Music}
  \city{Berlin}
  \country{Germany}
}
\author{Alexander Buchholz}
\authornotemark[1]
\email{buchhola@amazon.de}
\orcid{0000-0001-6521-7583}
\affiliation{%
  \institution{Amazon Web Services}
  \city{Berlin}
  \country{Germany}
}
\author{Pola Schw\"obel}
\authornotemark[1]
\email{schwobel@amazon.de}
\orcid{0000-0003-4846-1917}
\affiliation{%
  \institution{Amazon Web Services}
  \city{Berlin}
  \country{Germany}
}
% \author{Jan Malte Lichtenberg}
% \affiliation{%

% \affiliation{%

% \affiliation{%

% \affiliation{%

\renewcommand{\shortauthors}{Lichtenberg, Buchholz, Schwöbel}

\begin{abstract}
The issue of popularity bias---where popular items are disproportionately recommended, overshadowing less popular but potentially relevant items---remains a significant challenge in  recommender systems. Recent advancements have seen the integration of general-purpose Large Language Models (LLMs) into the architecture of such systems. This integration raises concerns that it might exacerbate popularity bias, given that the LLM’s training data is likely dominated by popular items. However, it simultaneously presents a novel opportunity to address the bias via prompt tuning. Our study explores this dichotomy, examining whether LLMs contribute to or can alleviate popularity bias in recommender systems. We introduce a principled way to measure popularity bias by discussing existing metrics and proposing a novel metric that fulfills a series of desiderata. Based on our new metric, we compare a simple LLM-based recommender to traditional recommender systems on a movie recommendation task.
We find that the LLM recommender exhibits less popularity bias, even without any explicit mitigation. 
\end{abstract}





\received{20 February 2007}
\received[revised]{12 March 2009}
\received[accepted]{5 June 2009}

\maketitle

\section{Introduction}
\pagestyle{empty}
Recently, general-purpose large language models (LLMs) have achieved astonishing successes across a wide range of tasks such as summarization, information extraction and content creation. In many domains, LLMs are used as foundation models---multipurpose tools that replace task-specific machine learning models.
% one such domain: recsys
The broad applicability of LLMs, coupled with their potential for use in intent-following or conversational recommender systems, has spurred interest in exploring their role in this domain~\cite{fan2023recommender,zhang2021language,luo2023recranker,gao2023policy,xu2024prompting}.
Our work analyzes LLMs as recommender systems along a specific dimension, popularity bias.

% what is popularity bias
% potential harms, tease on the idea of dynamics over time 
% ways of defining it
% normative commitmends: why we go with the "user-rec-comparison" approach

Popularity bias occurs when a recommender system disproportionally surfaces popular items (see Sections \ref{sec:background} and \ref{sec:quantifying_popularity_bias} for details), and has been present in recommender systems for decades \cite{bedi2014using,steck2011item}. It is still present today in applications such as modern streaming services that rely on recommender systems to guide the user’s consumption and exploration behavior. The potential negative impact of this bias ranges from filter bubbles over reduced user satisfaction, unfairness towards content producers to lost economic opportunity for platform providers. Strategies to quantify, explain, and mitigate popularity bias have seen growing interest lately, see \cite{klimashevskaia2023survey} for a recent survey. 

% maybe: pop bias in LLMs vs traditional recommender systems? 
Just like with standard recommender systems, popularity-biased behavior would not be unexpected for an LLM-based recommender: during training, the model has likely encountered popular content more often than lesser known content. Training data biases have been shown to propagate into model generations in other contexts such as geographical \cite{schwobel2023geographical,zhou2022richer} and gender-occupation biases \cite{rae2021scaling,liang2022holistic,huang2019reducing}.
On the other hand, LLMs also provide a new opportunity to mitigate popularity bias due to their natural language interface. We can simply \textit{ask} the LLM to recommend more niche content. Such self-debiasing \cite{schick2021self} has been proven effective across a range of bias-mitigation tasks  \cite{meade2021empirical}.

With the aim of quantifying the popularity bias in LLM-based recommender systems, we face a fundamental challenge: the literature lacks a consensus on what is an appropriate metric for measuring  popularity bias~\cite{klimashevskaia2023survey}. Therefore, our first contribution is the development of a principled framework for measuring popularity bias in recommender systems. We start by outlining a set of desiderata that a metric should satisfy, focusing on interpretability and statistical robustness. We then assess existing metrics against these desiderata and introduce a new metric that satisfies them.

To conduct our experiments, we propose a simple LLM-based recommender system, which can be built on top of any general-purpose LLM, and evaluate its popularity bias against a suite of baselines. We then mitigate the bias via prompting and study the resulting accuracy/popularity-bias trade-off.

\section{Background and Related Work} \label{sec:background}
\subsection{Traditional top-\textit{k} recommender systems} \label{sec:recsys}
Traditional top-\textit{k} recommender systems (RS) serve the purpose of selecting a set of $k$ most relevant items out of a much larger content pool. 
Common RS such as collaborative-filtering methods~\cite{wang2006unifying,ning2011slim,he2017neural} operate on \textit{item-level}. That is, these models learn an embedding space representing each candidate item and then recommend the \textit{k} unseen items that are closest in embedding space to the items previously consumed by the user (item-based collaborative filtering) or items consumed by similar users (user-based collaborative filtering). Different algorithms provide different ways of learning such embedding spaces but they are almost always trained on past user interaction data such as attributed clicks, purchases, or subscriptions.

Several research works have studied popularity bias in traditional recommender systems~\cite{chen2023bias,abdollahpouri2017controlling,zhao2013opinion,klimashevskaia2023survey}, and have investigated possible sources. Most arguments revolve around the idea that the historical interaction data used to train the recommender system contain, almost by definition, more behavioral signals from popular items than from unpopular ones. Collaborative filtering algorithms then naturally give more weight (on average) to the ``naturally more popular'' items, unless explicitly avoided by the algorithm~\cite{kamishima2014correcting,abdollahpouri2017controlling,saito2022fair}. Once such a recommender system is deployed, the popular items receive even more exposure and thus engagement. Consequently, the behavioral data that is gathered to train future generations of the recommender system will be even more popularity biased, resulting in a vicious circle of ever-increasing popularity bias~\cite{chen2023bias}.

\subsection{Large language models as top-\textit{k} recommender systems} \label{sec:llmrec}
LLM-based top-\textit{k} recommenders take a fundamentally different approach in that they are language-based and operate on a \textit{token-level}. In their simplest form, these models generate free-form text recommendations based on a prompt that describes the recommendation task and the user. The recommendations in text form are then parsed and resolved to the respective items in a catalogue before they are shown to the user. 

While such models are currently difficult to scale and use in practice due to their slow auto-regressive generation procedure and arguably have not yet reached state-of-the-art performance in traditional recommender evaluation protocols, they provide various advantages that open up entirely new possibilities for recommender systems:
First, LLM-based recommenders can be built on top of (open-source) general-purpose foundational models~\cite{zhang2021language,fan2023recommender}, reducing the need for behavioral data collection and model training.
    \footnote{This is particularly true for the movie domain because the training data of most current LLMs is based on the internet, which includes large amounts of movie reviews, discussions, and data bases.} 
    % Thus, there is less need for collecting vast amounts of behavioral training data and no need for model training, which could lead to a democratization of the recommender system landscape. %What if everyone could build their own recommender system? 
    % That being said, behavioral data could remain useful for fine-tuning the LLM recommender, and there are also approaches that train new models from scratch~\cite{cui2022m6}.
Second, LLM-based recommenders can generalize across different content types of a single domain (e.g., music, video, podcast in a media streaming service) or even across domains (e.g., the \textit{P5} model~\cite{geng2022recommendation}). Cross-domain generalization is notoriously difficult with feature-based recommender systems because of disjoint feature sets and varying consumption across domains.
Third, due to their natural-language interface, LLM-based recommenders enable novel paradigms including intent-based recommmenders~\cite{mehrotra2019jointly,benedict2023intent} that allow the user to formulate their complex intent in natural language (e.g., 
    \textit{``I want to watch a New York gangster movie featuring a female lead role''}, 
    \textit{``Movies that likely inspired the cinematographic style of Wes Anderson''}). In conversational recommenders~\cite{jannach2021survey}, the initial intent-based recommendations can additionally be refined by the user through follow-up requests. 

% add a general blob connecting this section and the next, like so: 
% While LLM-based recommenders promise these advantages, they might be prone to predicting popular items at overly high rates. As discussed in Section \ref{sec:recsys} this phenomenon, termed popularity bias, is well-known to exist also in standard recommender systems and might be exacerbated by the LLM-based modeling approach. 
% TODO: tidy up how and where we relate the pop bias of LLMs to the pop bias of standard RS
% As for standard RS, LLMs will be tilted towards what they have learned in their training set. Most current general-purpose LLMs are trained on a big chunk of the internet, which contains discussion, reviews, and ratings about the items to be recommended. For instance, in the case of a LLM-based movie recommender system, one would hope that numerous movie reviews and discussions in internet forums provide a good base for intent-based recommendations because people reveal why the they liked or disliked certain movies. However, it is also clear that, on average, there will be more of these discussions and reviews for more popular content. Consequently, a relatively unspecific query such as, for instance, ``\textit{Recommend an action movie.}'' is likely to recommend a highly popular action movie. 

However, as for standard RS, LLMs will be tilted towards what they have learned in their training set, and are prone to predicting popular items at overly high rates, resulting in popularity bias.

\subsection{Popularity bias of recommender systems} \label{sec:related_work_pop_bias}
The literature has produced an abundance of metrics formalizing different interpretations of popularity bias~\cite{chen2023bias,abdollahpouri2017controlling,zhao2013opinion,klimashevskaia2023survey, zhao2022popularity}. This diversity in metric selection, while reflective of varied study objectives and applications, poses a challenge in achieving generalizability across research findings, in particular because the normative commitments as to why a particular metric has been selected often remain unstated. 
% DONE!? \todo{Explain in more detail than in the introduction what popularity is, why it is harmful, what our basic assumptions are. In the end refer to Section \ref{sec:quantifying_popularity_bias} saying that we go into more detail about existing metrics and define our own.}
% [Klimashevskaia et al 23: ... a certain popularity bias in the underlying algorithms, which means that the algorithms may have a tendency to focus on already popular items in their recommendations.]
% [Klimashevskaia et al 23: ... meaning that they (RS) often focus on rather popular items in their recommendations..]
% [Klimashevskaia et al 23: A recommender system faces issues of popularity bias when the recommendations provided by the system focus on popular items to the extent that they limit the value of the system or create harm for some of the involved stakeholders.]
% [Abdollahpouri et al 2020 “popular items are recommended even more frequently than their popularity would warrant.”] 

Following Klimashevskaia et al.\cite{klimashevskaia2023survey}, we start with the general definition of \textit{popularity bias} as a property of a recommender system that is present ``\textit{when the recommendations provided by the system focus on popular items to the extent that they limit the value of the system or create harm for some of the involved stakeholders}''~\cite[p.8]{klimashevskaia2023survey}.  

Stakeholders include the service provider (e.g., a media streaming service or a job board), the content provider (e.g., artists, labels, companies that hire), and the user of the service (e.g. music listeners, job seekers). Popularity bias can have negative effects on all three stakeholders: the user might be bored by repeatedly receiving obvious or unoriginal ``mainstream'' recommendations; the service provider might suffer customer churn because of their decreased engagement or recommend only items that would have been sold anyways (and thus miss out on long-tail sales~\cite{anderson2006long}); and new or niche content providers might have a hard time entering or be driven out of the market due to limited exposure.
%
% From an ethical perspective, the latter is particularly problematic when it affects content creators from underrepresented demographic groups. 
Specifically, if a recommender system inordinately promotes items by a group of already popular artists, it may limit exposure of historically disadvantaged groups of content creators, hereby reinforcing this inequality \cite{deldjoo2021explaining}. This is a common concern in the fair ML literature: existing biases are picked up and exacerbated by ML-based systems \cite{barocas-hardt-narayanan,sheng2021societal}.  % ch.1 specifically
%In addition to this type of unfairness, popularity biased recommenders might violate normative claims of diversity and novelty. If we subscribe to diversity of the recommendation as a value in itself, a popularity biased recommender causes harm by reducing diversity. 

Note that while Klimashevskaia et al.'s definition of popularity bias does not prescribe a specific way of how the bias should be quantified, it 
excludes some existing definitions. % For example, Zhao et al. \cite{zhao2022popularity} focuses solely on unequal popularity distributions of the catalogue items, without reasoning about the the popularity of the recommendations.
For example, unlike in the popularity bias definition by Zhao et al. \cite{zhao2022popularity}, it is acceptable for some items to be more popular than others a priori. It is equally acceptable to recommend 
 popular items at higher rates; this would constitute popularity bias according to the definition by Abdollahpouri et al. \cite{abdollahpouri2017controlling} which is reminiscent of the fairness metric demographic parity \cite{calders2009building}. Instead, we argue that recommending more popular items more often is acceptable if they indeed are more \textit{relevant} to the user, following an interpretation similar to equality of opportunity \cite{hardt2016equality}. Predicting popular items starts to ``\textit{limit the value of the system or create harm for some of the involved stakeholders}'' when predicted popularities exceed the user's base popularity in expectation.
 %The popularity of a recommendation for any user should match the user's base popularity. 
 We will formalize this in the following section.

\section{Quantifying popularity bias} \label{sec:quantifying_popularity_bias}

 In working towards reducing the ambiguity of existing popularity bias measures, we propose a framework for defining a popularity bias metric for a given problem and data set. More specifically, we first define a parametrized formulation of popularity bias that generalizes several existing popularity bias metrics for specific parameter settings. We then define a set of theoretical desiderata for an interpretable and statistically robust popularity bias metric and evaluate existing metrics against these standards. Finally, we introduce a new metric that satisfies our desiderata.

As motivated in Section \ref{sec:related_work_pop_bias}, we quantify popularity bias of a recommender system with respect to the user's experience. A recommender system is positively (negatively) popularity biased if it recommends popular items at higher (lower) rates than appropriate for that user. 

We establish the following basic assumptions on how popularity is measured: 
\begin{enumerate}
    \item \textbf{Data}. The data set provides raw popularity scores $\phi(a)$ for every item $a$. Raw scores are an aggregate consumption measure over all users (e.g., total number of reviews for a movie, or total number of plays for a song). 
    Online consumption patterns tend to have heavy tails, 
    i.e., a few items account for the major share of interactions. 
    Empirically, item popularity follows a power law (Pareto distribution) arising from "rich get richer" dynamics, \cite{ratkiewicz2010characterizing,avramova2009analysis,clauset2009power}. 
    Power laws (and more specifically Pareto-distributions) have probability density functions of the form $p(x) \propto x^{-\alpha}$, for $x \geq x_{min} > 0$, where $\propto$ denotes proportional up to a constant. The coefficient $\alpha$ determines the tail behavior of the distribution. Power laws $\alpha \leq 2$ do not have a finite mean and for $2 < \alpha \leq 3$ the variance is infinite (and consequently all higher moments).  
    Hence, any metric based on averages (like mean, standard deviation, variance, skew and kurtosis) are ill-defined, as empirical averages do not converge. 
    Consequently popularity bias metrics with potentially huge fluctuations will have little scientific value. Therefore, raw popularity bias scores are often transformed implicitly by a function $g(\popa): \mathbb{R} \rightarrow \mathbb{R}$ for all items $a$. For instance, $g$ normalizes popularity scores to a certain range, or removes heavy tails. These transformations are often done without explanation, a point that we address below. 
    \item \textbf{Recommender popularity}. The popularity of a top-\textit{k} recommendation is an aggregate popularity score of all items in the slate, 
    $$\gagpr = h\left(\{ g(\popa) \}_{\ita \in \profr}  \right),$$ 
    where $h : \mathbb{R}^k \mapsto \mathbb{R}$ is a function that maps scores $\{ g(\popa) \}_{\ita \in \profr}$ to a single value,\footnote{As an example consider the empirical average, that maps scores to one value.} $k=|\profr|$, and $\profr$ is the set of all items chosen by the recommender. 
    \item \textbf{User popularity}. A user's popularity preference is defined by the aggregate popularity score of all (or the last \textit{n}) items the user consumed in the past (e.g., watch history for a movie streaming service), $$\gagpu = h\left(\{ g(\popa) \}_{\ita \in \profu}  \right),$$
    with $h$ as above and $\profu$ denotes the set of items a user interacted with. 
\end{enumerate}

%These base assumptions provide the main ingredients for a user based popularity-bias metric. 
To define a complete metric, one therefore has to make choices on (a) how the recommendation popularity $\gagpr$ and user popularity preference $\gagpu$ are aggregated on the recommender and user level and how they are combined to calculate the actual bias metric (with a slight abuse of notation, we will denote such a function by $M(u, r)$), and (b) how (if at all) the raw popularity scores are transformed by a function $g(\popa): \mathbb{R} \rightarrow \mathbb{R}$ for all items $a$. Most popularity bias studies only focus on the definition of $M$ and $h$, while treating $g$ independently. We argue that the properties of a popularity bias evaluation framework depend jointly on $M$, $h$ and $g$. % and the data set (that contains raw popularity scores). 
%Therefore, $M$, $h$, and $g$ should always be chosen simultaneously.

To make principled choices of $M$, $h$, and $g$, we define a set of desiderata for our final metric. We then discuss existing metrics and define our own metric.

% \ab{todo: add \cite{braun2023metrics}, discusses different popularity bias metrics}

\subsection{Desiderata for a popularity bias metric}\label{sec:desiderata}
We define the following desiderata for a popularity bias metric $M(r, u)$ and raw popularity transformation function $g$.

\begin{enumerate}
    \item \textbf{Well behaved}. We want our metric to be \textit{well behaved} in a statistical sense, i.e., more data leads to more stable results. Our aggregation function $h$ and transformation $g$ must be chosen such that $
    \gagpu$ (and $\gagpr$ respectively) become more precise with more data (i.e., when $|\profu|, |\profu| \rightarrow \infty$). This is crucial for reproducible and coherent statements about the popularity bias of a system. Without this requirement resulting metrics are meaningless as they measure random fluctuation in the data. This desideratum ensures \textit{construct reliabilty} \cite{jacobs2021measurement}.
    % stable first and second moments such that a law of large numbers and a central limit theorem hold., since, as we will see, most notions of popularity bias are constructed using averages. 
    % We deem this important as we expect more precise estimates of popularity bias with more data. This requirement will have an impact on both the way we define the metric transformation $g$ and its aggregation $h$. 
    
    \item \textbf{Centered around $0$}. If $\gagpr = \gagpu$ then $\deltaru = 0$. If there is no difference in aggregate popularity between recommendations and baseline popularity for a user, the metric should map to 0. This requirement influences $M$, and how we compare the sets $\profu$ and $\profr$. % The metric in \eqref{eq:gap} satisfies this.
    
    \item \textbf{Anti-symmetry}. $\deltaru = - \deltaur$. Popularity bias can be “positive” or “negative”. While most of the literature is concerned with “positive” popularity bias, i.e., recommending too popular items, “negative” popularity bias can have detrimental effects on the relevance of recommended items. We treat “positive” and “negative” popularity bias equally. We want to discover and mitigate both directions in our experiments. This will again influence the choice of $M$. 

    \item \textbf{Sensitivity to the long tail}. Our metric must be robust to large popularity values while at the same time being sensitive. The same absolute difference for large popularity values matters less than the same difference for small popularity values. 
    
    We define two sets of recommendations $\overline{r}$ and $\underline{r}$ where the main difference is that $~\overline{\cdot}~$ has an overall higher popularity than then items in$~\underline{\cdot}~$. We keep fixed the set of user items $u$. 
    We then want that 
    $$
     |  M(\underline{r} \oplus\epsilon, {u} ) -  M(\underline{r}, {u})| > 
     |   M(\overline{r} \oplus\epsilon, {u}) -  M(\overline{r}, {u})|,   
    $$  
    % $$
    % M(g(\underline{r}) \oplus\epsilon, {u} ) \geq M(g(\overline{r}) \oplus\epsilon, {u} )
    %  % |  \Delta AP(\overline{r} \oplus\epsilon, {u}) - \Delta AP(\overline{r}, {u})|,   
    % $$  
    where $\oplus$ denotes an increase of the popularity of one item by $\epsilon > 0$. 
    A shift in popularity for high popularity items of a recommender is relatively smaller than a shift in popularity of less popular items. This will influence all choices of $M$, $h$ and $g$. 
    This is important as a small change in the tail of the less popular items should have a relatively stronger influence than an increase in popularity of the most popular items. 
    % The metric in \eqref{eq:gap} is linear and as such does not satisfies this. 

    \item \textbf{Componentwise monotonicity}. We want a metric monotonic in the popularity scores. If popularity scores are increased by $\epsilon > 0$ for the recommender this should lead to the metric becoming larger (respectively smaller for an increase on the user side). 
    This requirements reads as 
     $$
         M(r \oplus\epsilon, {u}) -  M(r, {u}) > 0,  
    $$  
    using the same notation as before. This is important as the overall \textit{level} of popularity matters, which is assured by monotonic behavior. 
\end{enumerate}
% In particular, this means that recommending only highly popular or only long-tail items is not necessarily a bad thing as long as this does not conflict with the user's usual consumption patterns. 

% We next discuss the properties of existing popularity bias metrics.

\subsection{Existing metrics and our desiderata} \label{sec:existing_metrics}
We review a subset of existing metrics commonly used in the literature and discuss them in light of the above desiderata. We selected metrics that are consistent with our previously stated base assumptions. In particular, they can all be formalized within the same framework for specific functions $h$, $g$, and $M$, as shown in Table \ref{tab:comparison_metrics}. While this list of metrics is by no means exhaustive, we think that the selected metrics are representative of the main ideas and assumptions underlying most existing metrics in the literature. Table \ref{tab:comparison_metrics} provides a condensed overview of the alignment between these metrics and our desiderata. %For the sake of brevity we only briefly discuss the metrics and refer the reader to the respective publications for more details.
% Note, that we present a perspective using a comparison between a single user profile and a single recommender profile. Often popularity bias metrics are computed over user groups, which is achieved by averaging the resulting user level metric over groups. For the sake of simplicity we refrain from this step here. 

\begin{center}
\begin{table*}[t]
\small
 %\resizebox{\textwidth}{!}{
\begin{tabular}{ |l|  c | c | c | c | c |} 
 \hline
 Desiderata& \makecell{(Group) Average \\ popularity lift \cite{abdollahpouri2020connection}} & \makecell{Gini \\ index \cite{adomavicius2011improving,braun2023metrics}} & \makecell{Popularity\\ rank \\ correlation \cite{zhu2021popularity}}& \makecell{Herfindahl-\\Diversity \cite{adomavicius2011improving}} & \makecell{Log popularity \\ difference (ours) }\\
 \hline \hline
 
 (1) well behaved & \xmark & \cmark & \cmark & \cmark & \cmark \\ 
 (2) zero centered & \cmark & \cmark & \xmark & \cmark & \cmark\\ 
 (3) anti symmetric & \xmark & \cmark & \xmark & \cmark & \cmark \\ 
 (4) \makecell{long tail \\ sensitivity} & \xmark & \cmark & \xmark & \xmark & \cmark\\ 
 (5) monotonicity & \cmark & \xmark & \xmark & \xmark & \cmark\\ 
 \hline \hline
 Aggregation $h$ & mean & Gini coefficient & identity & sum of squares & mean \\ 
 Transformation $g$ & identity &  \makecell{normalization by \\ total popularity} & rank &  \makecell{normalization by \\ total popularity} & log \\ 
 Comparison $M$& relative difference & difference & correlation & difference & difference \\ 
\hline
\end{tabular}%}
\caption{Overview of common popularity bias metrics. We choose the aggregation function $M$ (amongst $\Delta, \Delta\%$ and $\%$) such that most desiderata are satisfied, indicated by \cmark. }
\label{tab:comparison_metrics}
\end{table*}
\end{center}

\paragraph{Average popularity lift}
Abdollahpouri et al.~\cite{abdollahpouri2020connection} use the lift of group average popularity as a metric to quantify popularity bias. Their metric is based on groups of users. For the sake of simplicity, we focus on overall average popularity (i.e., assuming only one single group). 
Average popularity at the user level $u$ (respectively, at the recommender level $r$) is defined as 
$$
\gapu = \frac{\sum_{a \in \profu} \popa}{|\profu|}, \hspace{1cm} \gapr = \frac{\sum_{a \in \profr} \popa}{|\profr|}.
$$
Importantly, no transformation of the raw scores is performed, that is, $g$ is just the identity and the score aggregation function is given by the empirical mean. The final metric, average popularity lift, is then given by 
\begin{equation} \label{eq:gap}
    \deltgaru = \frac{\gapr - \gapu}{\gapu}. 
\end{equation}

This metric, unfortunately, is at odds with several of our desiderata. 
Regarding Desideratum (1), note that the above metric is given as an empirical average of the samples of a random variable, that is, the raw popularity scores. 
Therefore, we would expect that the average converges to its expectation, i.e., $\gapu \rightarrow \mathbf{E}{\popa}$ as the number of samples $|\profu| = n \rightarrow \infty$. 
%The conditions for such a law of large numbers to hold are rather weak, i.e., $\mathbf{E}{\popa}$ and $\mathbf{E}{\popa^2}$ must be both finite (to give a simple version of the law of large numbers). Additionally, if the variance is infinite, no central limit theorem holds, which makes the construction of confidence intervals difficult. 
%However, it turns out that average popularity bias, is a poorly defined metric. 
However, if raw popularity scores follow a power law, as we argue at the top of Section \ref{sec:quantifying_popularity_bias}, this might not be the case; the metric might not converge to anything meaningful even with an infinite amount of data. The same reasoning is also valid for metrics that are derived from some form of average popularity, such as those proposed in 
\cite{deldjoo2021explaining} and \cite{lesota2021analyzing}, which use a popularity bias metric based on empirical moments of the raw scores. 

Furthermore, the metric in \eqref{eq:gap} also does not satisfy the desideratum  on anti-symmetry. 
Consider a user with $\gapu = 0.5$ and a recommendation with $\gapr = 1$ Then $\deltgaru = 1$ but $\deltgaur = -0.5$. In other words, \eqref{eq:gap} reports relative change (popularity bias in this case is 1, or 100\%, because the recommendation popularity is 100\% larger than the user base popularity).
Although this could be fixed by omitting the numerator, this would only fix the required anti-symmetry, not the other desiderata. 
Finally, the metric in \eqref{eq:gap} also weighs changes of the popularity equally, regardless of the scale, putting it at odds with our required long tail sensitivity. On the other hand the metric is monotonic, satisfying Desideratum (5).

\paragraph{Gini index}
The Gini index \cite{gini1921measurement} is a global measure of inequality of a distribution. For the case of popularity, we define it by considering the set of items consumed by a user and the items surfaced by a recommender as 
\begin{eqnarray*}
    Gini(r) =  \sum_{i =1}^{ |\profr|} \left( \frac{2i - |\profr| -1}{|\profr|} \right) \left( \frac{\popai}{\sum_{a_i \in \profr} \popai} \right), \\ Gini(u) = \sum_{i =1}^{ |\profu|} \left( \frac{2i - |\profu| -1}{|\profu|} \right) \left( \frac{\popai}{\sum_{a_i \in \profu} \popai} \right). 
\end{eqnarray*}
For the specific purpose of this metric we assume that the items $\popai$ are sorted in increasing order (i.e., $\popai \leq \phi(a_{i+1})$), see also \cite{braun2023metrics,adomavicius2011improving}. 
As these measures are defined at the level of the recommender (respectively the user) we need some way of comparing these two values. 
We suggest to use either some form of (relative) difference or ratio:
\begin{eqnarray*}
    \Delta Gini(r, u) = Gini(r) - Gini(u), \\ \% \Delta Gini(r, u) = \frac{\Delta Gini(r, u)}{Gini(u)}, \\ \% Gini(r, u) = \frac{Gini(r)}{Gini(u)}.
\end{eqnarray*}
This comparison finally amounts to comparing the inequality of popularity in the recommender profile vs. the inequality of popularity in the user profile. 
Note that only the first metric $\Delta Gini(u,r)$ is anti-symmetric and zero centered, which fulfills our Desiderata (2) and (3). The metric $Gini(r)$ is well defined in case of finite variance and reasonable estimators exist in the case of infinite variance, see \cite{yitzhaki2013gini,fontanari2018gini}, satisfying Desideratum (1). 
A small change in a large popularity value has smaller impact than the change of the popularity value of a small value, which fulfills our requirement of the long tail sensitivity. However, as a measure of inequality within a distribution, the Gini index is, in general, not monotonic. Imagine a case of an equal popularity of all items. If now all popularity values of the recommender are lifted by an equal increment, the Gini index remains unchanged, as the overall inequality is not varied, which is at odds with Desideratum (5). 

%Note that most studies do not reason about the raw popularity score transformation function $g$ even though it can be crucial (in particular for sensitivity to the long tail).

\paragraph{Popularity rank correlation}
For the rank correlation of a popularity bias metric we use the definition of \cite{zhu2021popularity} and adjust it to our notation. The popularity rank correlation measures how strong the popularity measured by the popularity scores $\popa$ of a set of items is correlated with the rank decided by a recommender. 
We define PRU as 
$$
PRU(r, u) = SRC \left[ \{\popa \in \profu \cap \profr \}, \text{rank}_r (\{a \in \profu \cap \profr \}) \right],
$$
where $SRC$ denotes the Spearman-rank correlation, $\profu \cap \profr$  denotes the intersection of items that are both part of the recommender profile and the user profile and $\text{rank}_r(\cdot)$ is the function that assigns to every item $a$ a rank decide by the recommender $r$. Note that this definition requires a notion of ranking by the recommender which is not required by the other metrics. Additionally, this metric intersects items from the user and recommender profile, which puts it slightly at odds with our imposed framework.  
The metric is well-behaved as ranks can be computed without additional requirements over user and recommender profiles. However, the metric is neither zero centered, nor anti-symmetric. 
This metric does not satisfy the long tail sensitivity and monotonicity  desiderata as changes in the value of the popularity score have no influence as long as the rank is not changed.

\paragraph{Herfindhal index}
The Herfindhal-index \cite{herfindahl1997concentration} was originally developed as a measure of economic concentration and has since been applied to measure popularity dispersion, see \cite{adomavicius2011improving}. 
We define it at the level of a recommender (a user respectively) as
$$
    H(r) = \sum_{a \in \profr} \left( \frac{\popa}{\sum_{a \in \profr} \popa} \right)^2, ~~ H(u) = \sum_{a \in \profu} \left( \frac{\popa}{\sum_{a \in \profu} \popa} \right)^2.
$$
The metrics are again defined at the user and recommender level, therefore we need an approach for computing a measure of popularity bias based on the two. Analogous to above we identify the following form of (relative) difference or ratio:
$$
\Delta H(r, u) = H(r) - H(u), ~~ \% \Delta H(r, u) = \frac{\Delta H(r, u)}{H(u)}, ~~ \% H(r, u) = \frac{H(r)}{H(u)}.
$$ 
The metric normalizes individual popularity scores by the sum of all scores in the profile, making it well-behaved (the metric is bounded between 0 and 1).  
By choosing the difference of the values, we can satisfy Desiderata (2) and (3). 
However, large popularity values have an overall dominant effect (due to the squaring of the values), putting it at odds with Desideratum (4),  Desideratum (5) (monotonicity) is also not fulfilled.

% Suggestion:
% use a log transform of the random variables, remove the normalization. 
% log transform will both make the metric well defined and allows sensitivity to the long tail

% TODO: 
% define a new transformed metric $M_f(r,u)$ where $f$ is a function that transforms the raw scores to satisfy our desiderata. 

\subsection{Properties of the log popularity difference metric}
We suggest a new metric, denoted \textit{log popularity difference}, that satisfies all our desiderata. It is conceptually similar to the average popularity lift metric but drops its normalization term (to ensure anti-symmetry) and introduces a log transformation of the raw popularity values (to ensure statistical robustness). The log popularity difference metric is defined as
\begin{equation} \label{eq:improvedpb}
    \Delta M (r,u) = \frac{\sum_{\ita \in \profr} \log \popa}{|\profr|} - \frac{\sum_{\ita \in \profu} \log \popa}{|\profu|}.     
\end{equation}
For data with popularity scores $\phi(a)$ following a Pareto distribution, this metric satisfies all Desiderata:  
Desideratum (1): it is well-behaved because the log transformation of a Pareto-distributed random variable follows an exponential distribution with well-defined mean and variance; (2): it is centered around zero; (3): the metric is anti-symmetric; and (4): changes in the tail (for high popularity values) have a weaker impact than changes for small popularity values due to the log transformation. Finally, the metric satisfies Desideratum (5) because the log transformation is monotonic itself and thus preserves the monotonicity of the average popularity aggregation function. We will use this new metric to measure popularity bias in our experiments.

\section{Experiments}

We aim to answer the following research questions in our experiments:
\begin{itemize}
    \item (RQ1) How does a recommender based on off-the-shelf LLMs compare to traditional behavioral recommenders in terms of both recommendation accuracy and popularity bias ?
    \item (RQ2) How effective are simple prompt-based popularity bias mitigation strategies? How do these mitigation strategies trade off popularity bias against recommendation accuracy?

\end{itemize}

We conduct our experiments in the movie recommendation domain using the `MovieLens 10M` data set~\cite{harper2015movielens}. The data set contains 10 million ratings on 10,000 movies from 72,000 users. We chose this dataset for the following reasons: 1) we expect general-purpose LLMs to be fairly good off-the-shelf movie recommenders, given the vast amount of movie-related content on the internet including reviews and discussions; and 2) we use the \textit{LensKit}~\cite{ekstrand2020lenskit} library to evaluate standard recommendation algorithms on the task and compare popularity bias and relevance of the LLM recommender to such classical methods. We start by defining the LLM-based recommender.

\begin{figure*}
\centering
\includegraphics[width=0.7\textwidth]{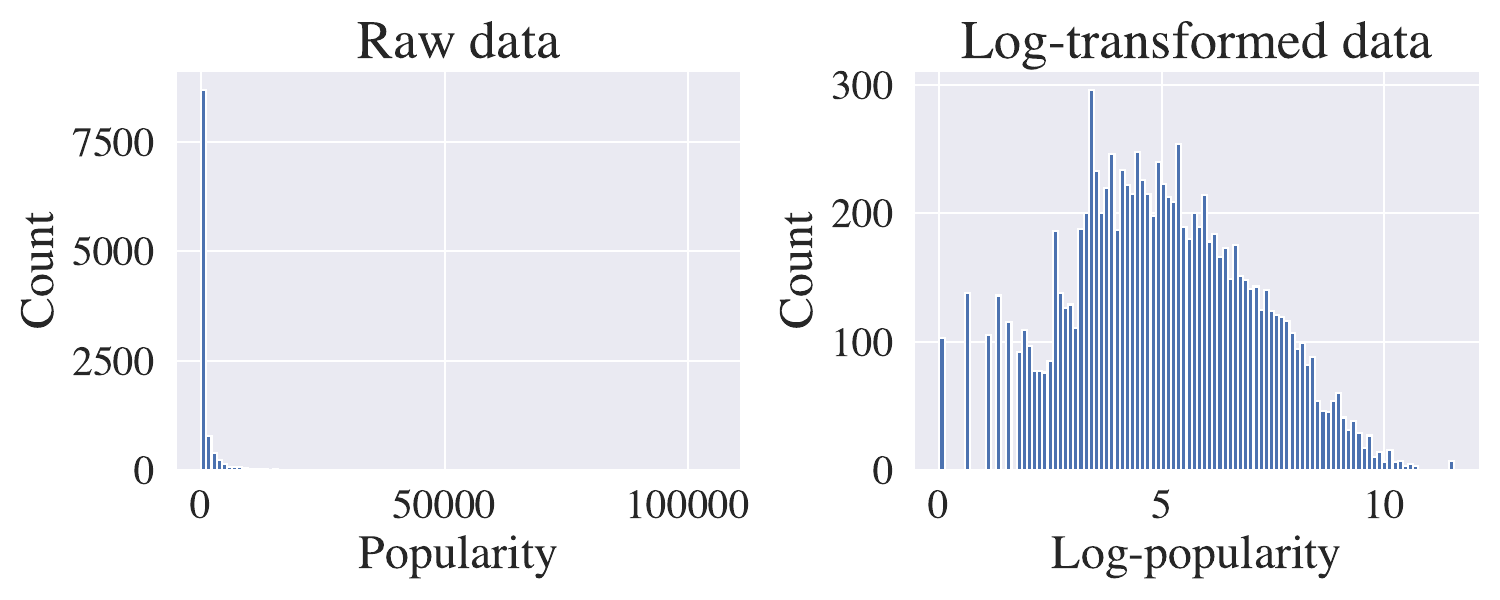} 
\caption{\textbf{Popularity scores of the MovieLens dataset.} Left are the raw scores, i.e., counts of how often a movie has been rated. We run a goodness-of-fit test (using the \href{https://erdogant.github.io/distfit/pages/html/index.html}{\textbf{distfit}} package) for several heavy tailed distributions and find that a Pareto-distribution, i.e., power law, is the best fit. The estimated coefficient is $\alpha =0.68$, makeing both mean and variance undefined. Right are log-transformed popularity scores. }
 \label{fig:ml_data}
\end{figure*}

% Old: The IMDb dataset contains popularity scores (i.e., the total number of votes a movie received) only. In this experiment we instead work on the MovieLens dataset which additionally contains behavioral features for the users. Specifically, we know which user rated which item. This allows us to extend our experiments in two directions. Firstly, we have ground truth available: As is standard in the literature, we consider a recommendation ``correct'' for a user if the user indeed has consumed this item. Secondly, this information allows us to evaluate standard recommendation algorithms on the task and compare popularity bias of the LLM recommender to such classical methods. 

\subsection{WOrld Knowledge recommender (WOK): A simple LLM-based movie recommender}\label{sec:wok}

% Recall that our goal here is to study popularity bias in recommenders using general-purpose LLMs to evaluate their suitability as off-the-shelf recommenders. 
% In particular, our goal is \textit{not} to construct a state-of-the-art recommender in terms of recommendation accuracy. 

We use a pre-trained LLM with a prompt template that simply asks to recommend a list of ten movies based on the watch history of a given user. The prompt template also contains formatting instructions so that we can reliably parse the natural-language output returned by the LLM and resolve it to a list of movie identifiers present in the MovieLens 10M dataset. This is needed to calculate relevance and popularity bias of the recommendation. Furthermore, the LLMs are  instructed to not recommend movies released after 2008, as these movies are not contained in the MovieLens data set, as well as movies that a user has already watched. % The exact prompt template is provided in the Appendix A.\ref{code:llm-prompt}.  

We call the resulting model \textit{WOrld Knowledge recommender} (WOK) because it relies entirely on the world knowledge (also known as parametric knowledge) acquired during pre-training, i.e., it has not explicitly been trained on the MovieLens 10M dataset to recommend movies. We use various off-the-shelf LLM-APIs from Anthropic and OpenAI as back-end LLMs and call the resulting models WOK-\textit{model-name} for different specific models  during our experiments. The Appendix contains further details on API parameter choices.

\subsection{Baselines}
In order to study the relationship between predictive performance and popularity bias we chose a range of simple baselines to compare our WOK recommender with. % which we recall in the following. 
%
% \subsection{IMDb Data}
% \subsection{Movielens Data}
% The IMDb dataset contains popularity scores (i.e., the total number of votes a movie received) only. In this experiment we instead work on the MovieLens dataset which additionally contains behavioral features for the users. Specifically, we know which user rated which item. This allows us to extend our experiments in two directions. Firstly, we have ground truth available: As is standard in the literature, we consider a recommendation ``correct'' for a user if the user indeed has consumed this item. Secondly, this information allows us to evaluate standard recommendation algorithms on the task and compare popularity bias of the LLM recommender to such classical methods. 
%
 % Removing the detailed description of the baselines since it would need a bit of work
% We define the user-item matrix as $X = \{x_{i,j}\}_{i,j}$ that is of dimension $K 
% \times M$, where $K$ is the number of users in our data set and $M$ is the number of items in the data. The entries $x_{i,j} \in \{0,1\}$ indicate if user $i$ has rated item $j$.
% Thus, we can rewrite the user-item matrix as $X = [v_1, \cdots v_K]^T$, i.e., each row of the matrix $v_i$ indicates the voting of the user on all items. Equivalently, we can write $X = [w_1, \cdots, w_M]$, where column $w_j$ contains all the rankings on item $j$. 
\textit{ItemKNN:} A traditional collaborative filtering algorithm, K-nearest neighbors based on item similarities. As recommended in the literature \cite{deshpande2004item} we pick $K=30$, i.e., we consider the $30$ most similar items for each item. 
    % \item \textit{ItemKNN:} Collaborative filtering algorithm based on item similarities.  Cosine-similarities between item $i$ and $j$ are computed as
    % \begin{equation}
    %    s_{ij} = \frac{\langle w_i, w_j \rangle}{\| w_i\| \| w_j\|}.
    % \end{equation}
    % $w_i$ is the binary vector indicating whether a user rated item $i$. This binary data is sometimes called \textit{implicit} data as opposed to the \textit{explicit} numerical ratings assigned by each user. 
    % % \ps{TODO: Better notation using $\Phi$ as the matrix of all ratings.} 
    % After computing the similarity, the $K$ largest values (i.e., closest neighbors) for each item are preserved, all other set to $0$. Hence we obtain a similarity matrix $S$ such that each the column $s_j$ contains the $K$ largest similarities for this item. As recommended in the literature \cite{deshpande2004item} we pick $K=30$. To make predictions, we are given a user's history $u$ (a one-hot encoding of whether or not the user has rated element $i$) and the number of items $N=10$ to be recommended. Then we multiply $S u = x$, $x$ containing the sum of all similarities for the items the user has watched. The items hat the user has already watched are set to $0$. We then recommend the $N$ items with the largest values in $x$. 
\textit{UserKNN:} Traditional collaborative filtering based on K-nearest neighbors based on user similarities \cite{shardanand1995social,wang2006unifying}. As before we pick $K=30$, i.e., we consider the $30$ most similar users for each test user. 
%     \item \textit{UserKNN:} Collaborative filtering algorithm based on user similarities. Again, cosine-similarities are computed, this time between the user vectors $u_i$,
% \begin{equation}
%     s_{ij} = \frac{\langle v_i, v_j \rangle}{\| v_i\| \| v_j\|}.
% \end{equation}
% Similar to before, we consider the $K=30$ largest values (i.e., closest neighbors) for each user in the user-user similarity matrix $S$.  ... \cite{wang2006unifying, shardanand1995social}.
\textit{Top Popularity Recommender:} Always recommend the $k$ most popular items.
\textit{Random Recommender:} Recommend $k$ items uniformly at random. %, $r = \{a_i\}_i^k, \ a_i \sim \mathcal{D}_{\{\text{uni}\}}$ with densities $p_{\text{uni}}(a_i) \propto 1$.

% \subsection{Sanity checks (properties of our evaluation framework/metric)}
% Toy experiments from 
% \url{https://quip-amazon.com/RpsJAa09R6yb/Popularity-Bias-in-Playlist-Generation-with-LLMs#temp:C:eaA6036d8f773754be8990cb0667}, could also go into the appendix. 

% \subsection{Preliminary Experiment: Hyperparameter Tuning for Baselines}
% It's surprising how much easier the prediction task seems to get when more than one movie are blocked out (see Fig.~\ref{fig:movielens_preliminary}). We will use `nr\_test\_users=1` as is common in the literature \cite{deshpande2004item, dacrema2019we}. (Note, however, that while this choice is somewhat common it is by no means the only possible choice, see \cite{dacrema2019we} for a discussion of this and the general reproducability issues withing RecSys research.)

% \begin{figure}
% \centering
% \includegraphics[width=0.5\textwidth]{images/nr_test_movies.pdf} 
% \caption{\textbf{Performance varies vastly with respect to the held-out test movies per user.}}
%  \label{fig:movielens_preliminary}
% \end{figure}

\subsection{Experimental Setup} \label{sec:experimental_setup}

We follow a standard evaluation protocol, where the ratings of each user are split into a training and testing set. The movies in the training set are used to train the algorithm, or in the case of the LLM recommender, as an input to the model at run-time (see Section \ref{sec:wok}). The recommender then generates a slate of recommendations. To measure the recommendation accuracy we report top five hit rate (HR@5) and top ten hit rate (HR@10). % which are computed as follows. For users in the test set, we leave out an individual movie. The goal is to predict this test movie from a user's previously watched items. We measure whether the test movie is among the first five (HR@5) or ten (HR@10) recommended movies.

To calculate the popularity bias of different algorithms, we approximate the raw popularity score of a movie by the total number of ratings it received. The raw as well as log-normalized scores are plotted in Fig.~\ref{fig:ml_data}. For LLM-based recommenders, we additionally report the number of invalid recommendations. Invalid recommendations occur when the LLM violates the instructions it has received via the prompt template. This could mean that the recommended item is formatted incorrectly, has already been watched by the same user in the past, is not referring to an existing movie title, or is a movie title that does indeed exist but is not part of the MovieLens data set. Note that the baseline recommenders have access to a valid candidate set and thus will always recommend valid titles. 

We repeat the experiment over five folds and report the average $\pm$ the standard error of the mean in Table \ref{fig:ml_results_fig}. Each fold consists of $1000$ users (i.e., we subsample the original dataset). 

\subsection{Results}
\begin{figure*}[t] 
    \begin{minipage}[c]{4.5cm}
    \includegraphics[width=\textwidth]{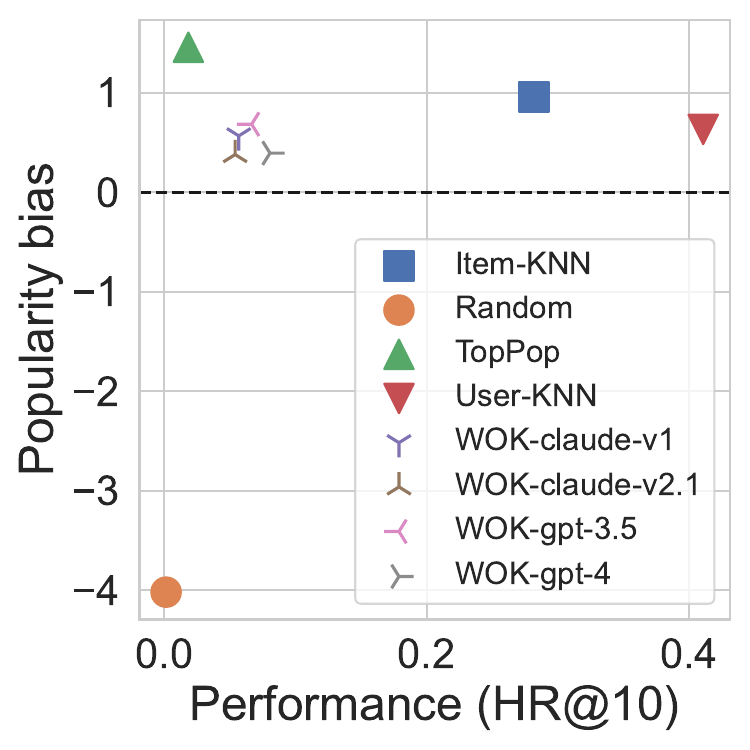} 
    \end{minipage}
    \hfill
  \resizebox{0.65\textwidth}{!}{ 
    \begin{tabular}{|c||c|c|c|c|} \hline 
                   &HR\@5 $\uparrow$      &  HR\@10   $\uparrow$ &  popularity bias $\downarrow$ & unmatched $\downarrow$ \\ \hline \hline
        Random     & $0.001 \pm 0.001$ & $0.001 \pm 0.001$ & $-4.020 \pm 0.013$   & $0$         \\ \hline
        TopPop     & $0.005 \pm 0.001$ & $0.014 \pm 0.002$ & $1.455 \pm 0.010$    & $0$         \\ \hline
        ItemKNN    & $0.204 \pm 0.006$ & $0.282 \pm 0.006$ & $0.957 \pm 0.009$    & $0$         \\ \hline 
        UserKNN    & $\textbf{0.313} \pm \textbf{0.007}$ & $\textbf{0.411} \pm \textbf{0.007}$ & $0.630 \pm 0.010$    & $0$         \\ \hline 
        WOK-claude-v1  & $0.044 \pm 0.003$ &  $0.057 \pm 0.003 $  & $0.565 \pm 0.010$              & $1.360 \pm 0.016$    \\ \hline
        WOK-claude-v2.1 & $0.042 \pm 0.003$ & $0.054 \pm 0.003$ &  $\textbf{0.377} \pm \textbf{0.010}$  & $1.361 \pm 0.018$ \\ \hline
        WOK-gpt-3.5 &  $0.047 \pm 0.003$     &    $0.067 \pm 0.004$     &    $0.682 \pm 0.010$   & $0.613 \pm 0.013$          \\ \hline
        WOK-gpt-4 &  $0.050 \pm 0.003$     &    $0.081 \pm 0.004$     &        $0.392 \pm 0.008$          &       $0.869 \pm 0.013$         \\ \hline
    \end{tabular}
    } 
    \captionlistentry[table]{A table beside a figure} 
    \captionsetup{labelformat=andtable} 
    \caption{Results on a subsample of the MovieLens 10M dataset. We repeat the experiments $5$ times, with $1000$ users in each fold. Reported are the mean plus/minus one standard error of the mean.}  \label{fig:ml_results_fig}
  \end{figure*} 
Figure \ref{fig:ml_results_fig} and Table \ref{fig:ml_results_fig}  show the results. A perfect recommender would achieve HR@5=HR@10=1 while exhibiting zero popularity bias. Regarding predictive performance, the traditional user-based collaborative filtering algorithm (UserKNN) works best. It is also among the least popularity biased models. 

The lowest popularity bias, surprisingly, is achieved by the Anthropic Claude-based WOK model (WOK-claude-v2.1). In fact, only WOK-gpt-3.5 exhibits a higher popularity bias than the least-biased baseline model. Note also that the Random recommender has a strong negative popularity bias. This indicates that the watch histories of users contained in the MovieLens data set contain movies with much higher-than-average  popularity scores. 

We also note the low number of invalid items for the WOK models, with all models returning valid movie recommendations in at least $8.5$ out of $10$ cases. This means that the LLM recommenders not only correctly format the recommended movies, but also respect the query, that is, they recommend movies from valid years and not do not recommend movies that the user has already watched. 

The gap in predictive performance of WOK models compared to UserKNN and ItemKNN can likely be explained by the additional information that those models have access to. While WOK models in this current implementation only access an individual test user's watch history, the collaborative filtering baselines have access to the full history of all user-item interactions. One could make this additional information available to the WOK models by fine-tuning them or infusing the user-item interactions via Retrieval-Augmented Generation (RAG, \cite{lewis2020retrieval}).

\subsection{Popularity bias mitigation and minimization via prompting}
A range of mitigation strategies have recently been explored for different types of biases in LLMs. Interventions target the training data \cite{zmigrod2019counterfactual,zhou2022richer}, the learned representations \cite{bolukbasi2016man,ravfogel2020null} or employ fine-tuning \cite{schwobel2023geographical}. However, a surprisingly simple approach has been shown effective \cite{meade2021empirical}. The model can be asked to `self-debias' \cite{schick2021self} via its prompt. In our case, we attempt to leverage the model's world knowledge regarding whether or not a movie is popular, and then ask it to focus on less popular items for debiasing. If successful, this mitigation strategy would be greatly beneficial from a usability perspective: Unlike other methods, it allows a user to configure the desired level of popularity via a natural language interface -- no ML expertise required. 

\textbf{Mitigation.} To evaluate this method, we add the additional instruction ``\textit{Recommend movies that match the average popularity level of the movies the user watched in the past.
For instance, if the user mostly watched blockbusters, you should recommend movies that are also blockbusters.
If, on the other hand, the user watched less well-known movies, you should recommend niche movies.
}'' to each of the WOK recommenders. Every such recommender is marked by a \textit{-mitigate} suffix.

% first experiment with modifying the prompt in one of two ways:
% \begin{itemize}
%     \item ``\textbf{mitigate}''. The aim here is to bring popularity bias as close as possible to zero. Specifically, we add the \textit{-mitigate} suffix to every WOK model whose prompt contains the additional instruction: ``\textit{Recommend movies that match the average popularity level of the movies the user watched in the past.
% For instance, if the user mostly watched blockbusters, you should recommend movies that are also blockbusters.
% If, on the other hand, the user watched less well-known movies, you should recommend niche movies.
% }.''
% \item ``\textbf{minimize}''. The aim here is to reduce popularity bias as much as possible, even if this results in negative popularity bias. Specifically, we add the \textit{-minimize} suffix to every WOK model whose prompt contains the additional instruction: ``\textit{Recommend indie, niche, or less well-known movies, avoiding mainstream blockbusters.}''
% \end{itemize}

 \begin{figure*}[h!] 
    \begin{minipage}[c]{10cm}
    \includegraphics[width=\textwidth]{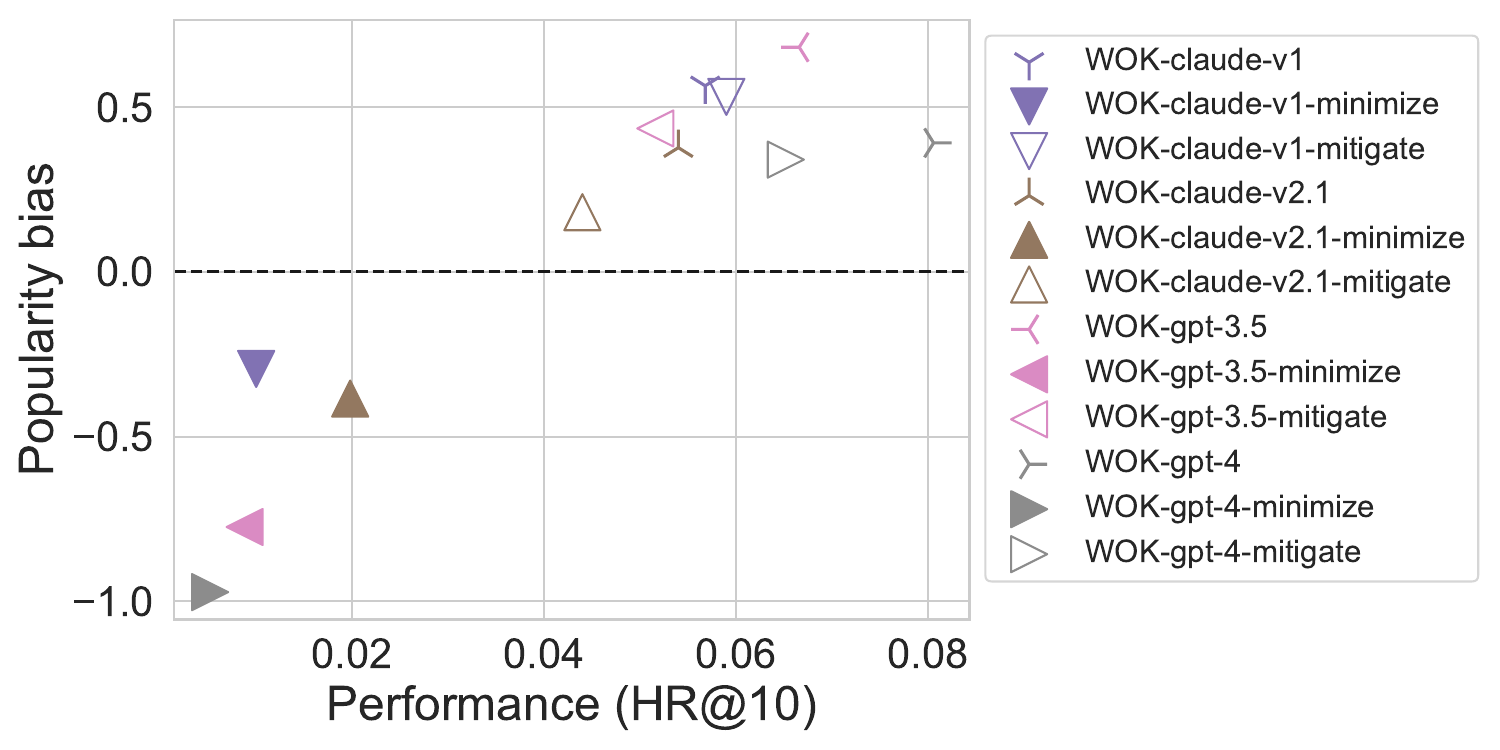} 
    \end{minipage}
    \hfill
  \resizebox{0.7\textwidth}{!}{%
    \begin{tabular}{|c||c|c|c|c|} \hline 
                   &HR@5 $\uparrow $     &  HR@10   $\uparrow$ &  popularity bias $\downarrow$ & unmatched $\downarrow$ \\ \hline \hline
        WOK-claude-v1  & $0.044 \pm 0.003$ &  $0.057 \pm 0.003 $  & $0.565 \pm 0.010$              & $1.360 \pm 0.016$    \\ % \hline
        WOK-claude-v1-minimize  & $0.008 \pm 0.001$ &  $0.010 \pm 0.001 $  & $-0.295 \pm 0.009$              & $2.001 \pm 0.031$    \\ 
        WOK-claude-v1-mitigate  & $0.046 \pm 0.003$ &  $0.059 \pm 0.003 $  & $0.532 \pm 0.010$              & $1.118 \pm 0.014$    \\ % \hline
        \hline
        WOK-claude-v2.1 & $0.042 \pm 0.003$ & $0.054 \pm 0.003$ &  $0.377 \pm 0.000$  & $1.361 \pm 0.018$ \\ % \hline
        WOK-claude-v2.1-minimize & $0.015 \pm 0.002$ & $0.020 \pm 0.002$ &  $-0.385 \pm 0.010$  & $0.787 \pm 0.016$ \\ 
        WOK-claude-v2.1-mitigate & $0.033 \pm 0.003$ & $0.044 \pm 0.003$ &  \textbf{0.181} $\pm \textbf{0.010}$  & $0.792 \pm 0.012$ \\ % \hline
        
        \hline
        WOK-gpt-3.5 &  $0.047 \pm 0.003$     &    $0.067 \pm 0.004$     &    $0.682 \pm 0.010$   & \textbf{0.613} $\pm$ \textbf{0.013}          \\ % \hline
        WOK-gpt-3.5-minimize  & $0.007 \pm 0.001$ &  $0.009 \pm 0.001 $  & $-0.775 \pm 0.013$              & $3.393 \pm 0.031$    \\ 
        WOK-gpt-3.5-mitigate &  $0.037 \pm 0.003$     &    $0.052 \pm 0.003$     &    $0.438 \pm 0.012$   & $1.105 \pm 0.022$         \\ % \hline
        \hline
        WOK-gpt-4 &  \textbf{0.050} $\pm$ \textbf{0.003}     &    \textbf{0.081} $\pm$ \textbf{0.004}     &        $0.392 \pm 0.008$          &  $0.869 \pm 0.013$         \\ % \hline
        WOK-gpt-4-minimize & $0.004 \pm 0.001$     &   $0.005 \pm 0.001$     &  $-0.972 \pm 0.009$          &       $1.549 \pm 0.019$         \\
        WOK-gpt-4-mitigate &  $0.045 \pm 0.003$     &    $0.0652 \pm 0.003$     &        $0.341 \pm 0.007$          &       $0.914 \pm 0.012$ 
        \\ \hline
    \end{tabular}
    } 
    \captionlistentry[table]{A table beside a figure} 
    \captionsetup{labelformat=andtable} 
    \caption{Results for the mitigation experiment. The results are grouped by base LLM in the WOK model. Bold numbers indicate best performance across all models.}  \label{fig:mitigation_results_fig}
  \end{figure*}

We use the same experimental setup as described in Section \ref{sec:experimental_setup}. Table \ref{fig:mitigation_results_fig} and Figure \ref{fig:mitigation_results_fig} show the results. The popularity bias of all \textit{mitigate}-models (hollow triangles in Figure \ref{fig:mitigation_results_fig}) is slightly decreased when compared to the respective base models (``tripods'' in Figure \ref{fig:mitigation_results_fig}). For all base-LLMs apart from claude-v1, this reduction in bias is traded off against a reduction in prediction performance. 

Note with the exception of claude-v2.1, the popularity bias values remain relatively close to their base values, or in other words, the mitigation strategy is not highly effective. 
% This small mitigation effect could be explained as follows. A good recommender that aims to recommend movies that are similar to a given set will try to match  likely due to the fact that  be explained by the fact that counterparts with unmodified prompts. This can be explained by the fact that even the 
We therefore also try a second, more extreme mitigation strategy to study how far we can push the WOK recommender into providing long tail recommendations.

\textbf{Minimization.} To this aim, we replace the \textit{mitigate}-instruction by the following instruction: ``\textit{Recommend indie, niche, or less well-known movies, avoiding mainstream blockbusters.}''  The resulting models are marked by a \textit{-minimize} suffix. 

The results are again shown in Table \ref{fig:mitigation_results_fig} and Figure \ref{fig:mitigation_results_fig}. The \textit{-minimize} strategy does indeed reduce popularity bias to a point where the resulting models show strong negative popularity bias. However, this strategy also results in a strongly decreased recommendation accuracy across all LLMs.

% All WOK models lose in recommendation accuracy, albeit to different degrees. In particular, while both GPT models have higher HR@10 than the Claude models before the mitigation strategy is applied, Claude models outperform their counterparts after popularity bias mitigation. 

Note also that the number of unmatched items increases considerably when using the mitigation prompt (except for claude-v2.1), which highlights an important limitation of this experiment. Recall that the MovieLens data sets contains ``only'' 10,000 movies and that WOK recommenders do not have access to this list of valid movies. As a consequence, WOK models that aim to minimize popularity bias might be punished for recommending niche movies that do not appear in the MovieLens catalogue, but could nevertheless be relevant to the user.

\subsection{Correlation of popularity bias metrics} \label{sec:correlations}
\begin{figure}[h!] 
    \includegraphics[width=0.5\textwidth]{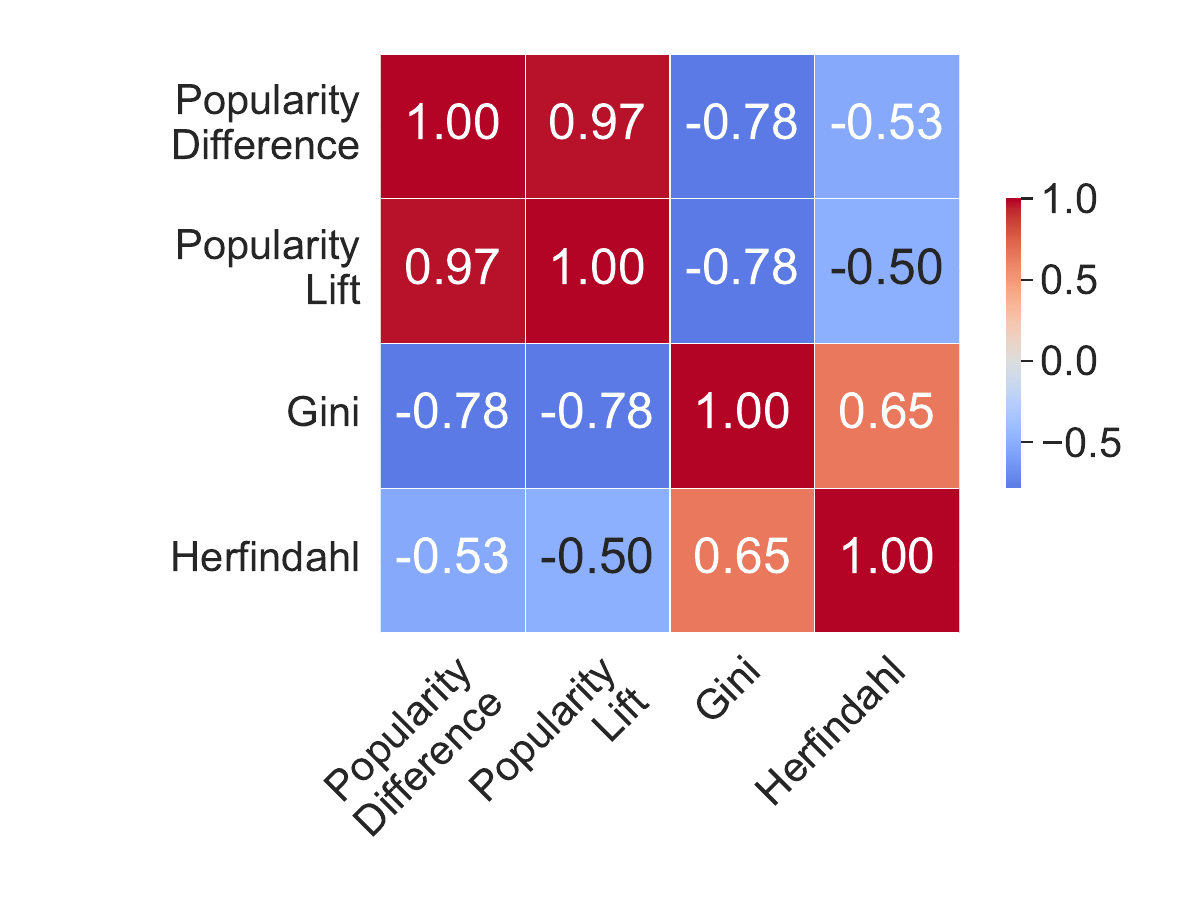} 
    \caption{Kendall’s Tau correlation coefficients between the various metrics measured across experiments reported in Tables \ref{fig:ml_results_fig} and  \ref{fig:mitigation_results_fig}.}
    \label{fig:correlation}
\end{figure}

We compare the metrics from Table \ref{tab:comparison_metrics} with each other based on the popularity bias values that are reported in Table \ref{fig:ml_results_fig} ($8$ values, original model performance for each metric) and Table \ref{fig:mitigation_results_fig} ($8$ additional values, model performance under two mitigation strategies for each metric). In total we base the correlation analysis on $8+8=16$ popularity bias values for four different metrics, i.e., Gini--diversity, Herfindahl--diversity, average popularity lift, and our suggested metric, the log popularity difference. Unsurprisingly, log popularity difference and average popularity lift show a very strong correlation, as measured by Kendal's $\tau$ \cite{kendall1938new}, as shown in Figure \ref{fig:correlation}. These two metrics are strongly correlated, as both metrics are very similar in nature, but average popularity lift does not satisfy the Desiderata (1), (3) and (4). For an analysis of the practical implications of this, see Appendix \ref{sec:appendix}.

Herfindahl-- and Gini--diversity measure fundamentally different things than popularity lift and log popularity difference (the difference of popularity \textit{diversity} between user and recommendation vs. the difference of \textit{average popularity} between user and recommendation). Hence, the two families of metrics are not expected to correlate well. In practice, we observe them to be negatively correlated in Figure \ref{fig:correlation}.

\section{Discussion and Further Work} \label{sec:discussion}

Recall that our goal was to study popularity bias in recommenders using general-purpose LLMs to evaluate their suitability as off-the-shelf recommenders, as opposed to constructing a state-of-the-art recommender in terms of recommendation accuracy.  We therefore deliberately kept the LLM recommender as light-weight as possible to ensure that our findings are reflective of the intrinsic bias in LLMs, rather than being confounded by complex, model-specific factors. In doing so, we aim to provide insights that are broadly applicable across various LLM implementations, contributing to a more generalized understanding of popularity bias in these systems. 
Nevertheless, the recommendation accuracy of the WOK recommender could likely be improved using various techniques, including 1) fine-tuning on a training set of successful recommendations; 2) ``grounding'' the recommendations in a movie catalog to avoid hallucinations (via RAG, \cite{lewis2020retrieval}); or 3) more advanced prompting techniques such as chain-of-thought prompting, or more advanced prompt optimization techniques~\cite{yang2023large,fernando2023promptbreeder}. Furthermore, these or similar automatic prompt optimization techniques could also be applied to find more balanced prompt-based mitigation strategies that find a happy medium between the two strategies we experimented with.

An open research question is the evaluation of intent-based performance of LLM-based recommenders. As discussed in Section \ref{sec:llmrec}, easily generating recommendations that are guided by user-intent is a compelling promise of LLM-based RS. In order to evaluate the performance in this task, one would require structured ground truth, or, as a starting point, use simple proxy tasks where the intent encodes known metadata such as director, genre, year or similar.
Similarly, while research in popularity bias is motivated by ``rich-get-richer'' effects over time, we have not modeled or measured temporal dynamics (see \cite{liu2018delayed,kannan2019downstream,zhang2020fair} for dynamical modeling approaches, and \cite{schwobel2022long} for an overview).

\section{Conclusion}

We have investigated the promise of LLM-based recommender systems focusing on a specific aspect of their performance and usability: popularity bias.
We started by constructing a measurement for the phenomenon. To do so, we have formulated desiderata for a popularity bias metric, and have evaluated existing metrics against these desiderata. Applying some adjustments to the Popularity Lift metric by Abdollahpouri et al.~\cite{abdollahpouri2019unfairness}, we have arrived at our metric: log popularity difference. We acknowledge that this metric may not suit all future studies on popularity bias given the topic’s diverse application domains and goals. However, we encourage future research to examine the assumptions and theoretical properties of their chosen metrics, with our framework potentially serving as a useful starting point.

Using our metric, we have compared traditional RS against the LLM-based models on the MovieLens 10M dataset. We have found the LLM-based recommenders to have moderate amounts of popularity bias; usually less than their traditional, collaborative filtering-based counterparts. 
In our mitigation experiment, we have found that it is possible to lower popularity bias further by including additional instructions in the prompt. In the extreme case, where the model is specifically instructed to `avoid mainstream blockbusters', we achieve negative popularity bias. This is accompanied by a drop in recommendation accuracy, which overall is lower for our naïve LLM-based recommenders compared to collaborative filtering baselines. 

% In our naïve implementation they have, however, performed much worse in terms of accuracy as measured by HR@5 and HR@10. While not the focus of this work, performance of LLM-based recommender systems could likely be improved dramatically, for example by employing the strategies in Section \ref{sec:discussion}.

We believe that LLM-based recommender systems will see widespread usage. We encourage practitioners to measure the popularity bias of such models, and, especially in light of its simplicity, experiment with popularity-debiasing via prompting before deploying such a system. % [+ sth with the downstream consequences and rich get richer effects]

\bibliographystyle{ACM-Reference-Format}
\bibliography{samples/references}

\appendix

\onecolumn
\section{Implementation details}

\subsection{Implementation details: LLM-based WOK recommender}

We used OpenAI's \texttt{gpt-3.5-turbo-0613} (``gpt-3.5'' in the main text) and \texttt{gpt-4-1106-preview} (``gpt-4'' in the main text) APIs, as well as Anthropic's \texttt{claude-instant-1.2} (``claude-v1'' in the main text) and \texttt{claude-2.1} (``claude-v2.1'' in the main text) APIs. All LLMs used a temperature parameter of $0.0$, a \textit{top-p} parameter of $1$  (default for both OpenAI and Anthropic APIs), and a \textit{top-k} parameter of $250$ (default). The base prompt template used for all experiments is given in Code Listing \ref{code:llm-prompt}.

\begin{codeblock}
\begin{lstlisting}[language=Python]
"""
You are a helpful movie-expert AI tasked with recommending a collection of movies based on a user's watch history. The user has watched the following movies in the past:
{watch_history}

# Output instructions
- Immediately start with the movies. Do not provide an introduction.
- Provide a list of {nr_items} movies.
- For each movie, start a new line, indicate the position in the movie list (that is, 1., 2., ...).
- Name the title of the movie (without quotation marks!) and then in parentheses the release year.
- Do not recommend movies that the user has already watched. Those are the ones listed above.
- Do not recommend movies that are newer than 2008.

Now create the movie list!"""
\end{lstlisting}
\caption{Prompt template used for the LLM movie recommender. The placeholder \texttt{watch\_history} is replaced by a list of movies watched by the user at runtime.}
\label{code:llm-prompt}
\end{codeblock}

\section{On the difference of average popularity lift and log popularity difference and the importance of statistical well-behavedness (Desideratum 1).} \label{sec:appendix}
% } 

\begin{figure*}[h!] 
    \includegraphics[width=0.5\textwidth]{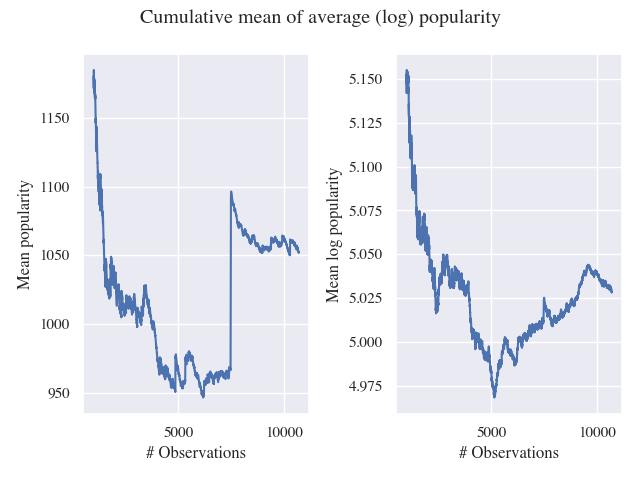} 
    \label{fig:popbias_convergence}
    \caption{Behavior of the average popularity bias as a function of the data points.} \label{fig:means}
  \end{figure*}

We have seen that average popularity lift and log popularity difference are strongly correlated (Section \ref{sec:correlations}), but differ in that average popularity lift does not satisfy the Desiderata (1), (3) and (4), see Table \ref{tab:comparison_metrics}. We have motivated the importance of these desiderata theoretically in Section \ref{sec:desiderata}, and investigate their practical implications here, specifically for well-behavedness (1). 
%Table \ref{TODO} shows the standard errors of the means under average popularity lift compared to log popularity difference. The standard error is one order of magnitude larger for the average popularity lift metric, 
Figure \ref{fig:means} illustrates the cumulative averages of popularities $\sum_{i+1}^N \phi(a_i)/N$ for $N$ a varying number of observations ($x$-axis) under average popularity lift (left) and log popularity difference (right). For the average popularity lift metric, there is a sudden jump when the largest popularity item is added (around $i=7500$). This results from the Power-law distribution of unnormalized scores ($g=$identity, Table \ref{tab:comparison_metrics}) and the resulting ill-behavedness. As a consequence, average popularity lift measurements will be highly sensitive to whether or not individual, high popularity items are included in recommendation or user sets $p_u$ or $p_r$, respectively, and will measure any differences in the low-mid popularity regime less reliably. 

\end{document}